\documentclass[12pt]{article}

\usepackage{amsmath,amsfonts}
\usepackage{amssymb}
\usepackage{hhline}
\usepackage{epsfig,cite}
\usepackage{stmaryrd}
\usepackage[usenames,dvips]{color}
\usepackage{fullpage}
\usepackage{verbatim}
   \allowdisplaybreaks
\makeatletter
\@addtoreset{equation}{section}
\makeatother

\newcommand{\be}{\begin{equation}}
\newcommand{\ee}{\end{equation}}
\newcommand{\bea}{\begin{eqnarray}}
\newcommand{\eea}{\end{eqnarray}}

\begin{document}

\thispagestyle{empty}

\begin{center}
\hfill CERN-PH-TH/2014-022\\
\hfill UAB-FT-752\\
\hfill EFI-14-2\\
\hfill  NSF-KITP-14-007
\begin{center}

\vspace{.5cm}

{\Large\sc General Focus Point in the MSSM}

\end{center}

\vspace{1.cm}

\textbf{ A.~Delgado$^{\,a}$, 
M.~Quiros$^{\,b}$ and C.~E.~M~Wagner$^{\,c}$}\\

\vspace{1.cm}
${}^a\!\!$ {\em {Department of Physics, University of Notre Dame\\ Notre Dame, IN 46556, USA}}

\vspace{.1cm}

${}^b\!\!$ {\em {Department of Physics, CERN-TH Division, CH-1211, Geneva 23, Switzerland and\\
Instituci\'o Catalana de Recerca i Estudis  
Avan\c{c}ats (ICREA)\\ Institut de F\'isica d'Altes Energies, Universitat Aut{\`o}noma de Barcelona\\
08193 Bellaterra, Barcelona, Spain}}

\vspace{.1cm}
${}^c$ {\em{Enrico Fermi Institute,  Department of Physics and Kavli Institute for \\ Cosmological Physics,
University of Chicago, Chicago, IL 60637, U.S.A. \\
HEP Division, Argonne National Laboratory, Argonne, IL 60439, U.S.A. \\
}}

\end{center}

\vspace{0.8cm}

\centerline{\bf Abstract}
\vspace{2 mm}
\begin{quote}\small
The minimal supersymmetric extension of the Standard Model (SM) is a well motivated scenario for physics beyond the SM, which allows
a perturbative description of the theory up to scales of the order of the Grand Unification scale, where gauge couplings unify.
The Higgs mass parameter is insensitive to the ultraviolet physics and is only sensitive to
the scale of soft supersymmetry breaking parameters.
Present collider bounds suggest that the characteristic values of these parameters may be
significantly larger than the weak scale. 
Large values of the soft breaking parameters, however, induce large radiative corrections to the Higgs
mass parameter and therefore the proper electroweak scale may only be obtained by a fine tuned cancellation
between the square of the holomorphic $\mu$ parameter and the  Higgs supersymmetry breaking square mass parameter. 
This can only be avoided if  there is a correlation between the scalar and gaugino mass parameters, such that 
the Higgs supersymmetry breaking parameter remains of the order 
of the weak scale. The scale at which this happens is dubbed as focus point.  In this article, we  define the general conditions required for this to happen, for different values of 
the messenger scale at which supersymmetry breaking is transmitted to the observable sector,  and for arbitrary 
boundary conditions of the sfermion, gaugino, and Higgs mass parameters.  Specific supersymmetry breaking scenarios in which these correlations may occur are also discussed.

 \end{quote}

\vfill

 \newpage
\section{Introduction}
The Standard Model (SM) of particle physics provides an excellent description of all known fundamental particle interactions, excluding gravity.  Mass generation in the SM is obtained via the Higgs mechanism, induced by the presence of a scalar (Higgs) field, which transforms in the fundamental representation of the SM~$SU(2)$ gauge group.  The properties of the recently discovered resonance at the LHC~\cite{ATLAS,CMS} seem to be close to the ones expected for  the SM Higgs boson, and hence the physics at the weak scale is well described by the SM. 

Supersymmetry (SUSY)~\cite{reviews}, 
as the simplest solution to the hierarchy problem, provides a well motivated extension of the SM, with new particle masses determined by the soft supersymmetry breaking mass scale.  If the sparticle masses are flavor independent, the supersymmetric particles effects tend to decouple in a fast way at scales below these masses and one recovers the SM as a low energy effective theory, as required by observations.  
The radiative corrections to the Higgs mass parameter are governed by  the soft  SUSY breaking scale, and therefore in order to avoid a large fine-tuning, the SUSY breaking masses should not be much larger than the weak scale~\cite{Barbieri:1987fn}.  More precisely, there is a specific combination of the SUSY breaking masses that determines the value of the low energy Higgs mass parameter and this combination should remain small (of the order of the weak scale) in order to avoid fine-tuning. 

In view of the increasingly stronger bounds set at the LHC on the masses of possible beyond the Standard Model (BSM) particles introduced to solve the hierarchy problem, as e.g.~the supersymmetric partners of the SM particles~\cite{ATLAS,CMS}, it is interesting to find regions in the parameter space where the fine-tuning problem is alleviated. In the particular case of  the minimal supersymmetric extension of the Standard Model (MSSM), an interesting observation was done in Ref.~\cite{Feng:1999zg}, in which it was demonstrated that, starting with universal supersymmetry breaking scalar masses at the GUT scale, the Higgs mass parameter could become small  based on the existence of a renormalization group equation (RGE) focus point (FP) at the electroweak (EW) scale $Q_{EW}$.  The original focus point allowed to obtain the proper electroweak symmetry breaking scale for large values of the squark and slepton masses and subleading values of the stop mixing parameter $A_t$ and gaugino masses $M_a$. More recently, this solution was reconsidered,  including the presence of large stop mixing $A_t$~\cite{Feng:2012jfa}, large (non universal) gaugino masses~\cite{Horton:2009ed,Antusch:2012gv,Kowalska:2014hza} and in the framework of gauge mediation~\cite{chino}.

In general, one would be interested in solutions in which the Higgs mass parameter does not scale with the rest of the soft supersymmetry breaking parameters.  This is relevant, since experimental data suggest that gluinos and sfermion masses may be much larger than the weak scale~\cite{ATLAS,CMS}.  If this were the  case, these particles would decouple at some scale $\mathcal Q_0 \gg \mathcal Q_{EW}$, and therefore the matching between the SM and the SUSY extension should be performed at the scale $\mathcal Q_0$, at which the heavy particles are decoupled. 
 
The matching condition yields a relationship between the SM Higgs boson potential parameters
\be
V(H)=-m^2|H|^2+\frac{\lambda}{2}|H|^4,
\ee
where $m^2(\mathcal Q_{EW})=\frac{1}{2}m_H^2$, and the supersymmetric parameters at the scale $\mathcal Q_0$ as~\cite{Delgado:2013gza}
\begin{align}
m^2&=\frac{m_{H_D}^2-m_{H_U}^2}{\tan^2\beta-1}-m_{H_U}^2-|\mu|^2
\label{eom}\\
\lambda&=\frac{1}{4}(g_1^2+g_2^2) \cos^2 2\beta+\frac{3h_t^2}{8\pi^2}X_t^2\left(1-\frac{X_t^2}{12} \right)
\label{lambdamatch}
\end{align}
where $X_t=\frac{(A_t-\mu/\tan\beta)}{m_Q}$, and $m_Q\simeq \mathcal Q_0$ is the stop mass parameter which we will consider as the generic value of the supersymmetric mass spectrum. As the $m^2$-parameter on the left-hand side of Eq.~(\ref{eom}) does run very slowly with the RGE scale $\mathcal Q$~\cite{Buttazzo:2013uya} we can replace it by $m_H^2/2$.

A heavy supersymmetric spectrum in general implies that the MSSM soft-breaking terms for the scalars and gauginos $(m_Q^2,\, m_U^2,\, m_{H_U}^2,\, M_a)$ are large at the high scale $M$ at which they are generated, in which case one expects also $m_{H_U}^2(\mathcal Q_0)$ to be large,  thus triggering a huge fine-tuning for Eq.~(\ref{eom}) to be satisfied. 
A rough definition of the sensitivity with respect to the model parameters $a$ was given in Ref.~\cite{Barbieri:1987fn} as
\be
\Delta_a=\left| \frac{\partial\log m_H^2}{\partial\log a} \right|
\ee
while the overall measure of fine tuning can be defined as $\Delta=\max_a \{\Delta_a\}$.
In particular the fine-tuning in $\mu$ is very special as its low energy value is determined by the equation of minimum (EOM)~(\ref{eom}) and it turns out that in the MSSM the fine-tuning is often dominated by $\Delta_\mu$\cite{Antusch:2012gv}. In the case of large or moderate values of $\tan\beta$ one gets from (\ref{eom})
\be
\Delta_{\mu^2}\simeq\left|1+2 \frac{m_{H_U}^2}{m_H^2}   \right|
\ee 
Therefore if there is a RGE FP at the scale $\mathcal Q_0$, which we define as $m_{H_U}^2(\mathcal Q_0)=0$, then in the moderate or large $\tan\beta$ regime $\Delta_\mu=1$ (i.e.~no fine-tuning), Eq.~(\ref{eom}) is widely insensitive to the boundary conditions at the scale $M$, and the fine-tuning is greatly alleviated. Of course there is an underlying tuning by which the hidden sector provides, at the scale of messenger masses, a particular pattern of values of the supersymmetry breaking (and $\mu$) parameters which will in turn predict a given FP of the RGE. In the absence of a precise theory of supersymmetry breaking this fine tuning in the underlying theory cannot be computed.

In this paper we have integrated the RGE from the high-scale $M$ to the low-scale $\mathcal Q_0$ and provided a formally analytical expression for $m_{H_U}^2(\mathcal Q_0)$ in terms of all supersymmetric parameters defined at $M$. For vanishing values of the hypercharge D-term, it turns out to be a linear combination of $m_Q^2$, $m_U^2$, $m_{H_U}^2$, $M_aM_b$, $M_a A_t$, and $A_t^2$, with coefficients that depend on $M$ and $Q_0$. We have made very accurate fits for the different coefficients fixing $\mathcal Q_0=2$~TeV and for $M \in [10^5,10^{17}]$ GeV so that the FP condition can be written as an algebraic equation involving $\log (M/\mathcal Q_0)$ and the soft-breaking terms at the scale $M$. The result, presented in Sec.~\ref{general}, is an easy-to-use equation from where the FP condition can be established for arbitrary boundary conditions and an arbitrary scale $M$. We also provide the (very simple) contributions that should be added for non-vanishing values of the hypercharge D-term. In Sec.~\ref{models} we have applied the general equations to some of the most popular models, including CMSSM and gravity mediated models, gauge mediated models and mirage models. We did not exhaust the different possible models (or made scatter plots on all models) as it should be trivial to apply our formulae to any particular model. Finally in Sec.~\ref{conclusion} we present our conclusions. Some technical details of  the calculation are postponed to App.~\ref{appendix}.

\section{General Focus Point}
\label{general}
We will assume that the MSSM soft-breaking terms, in particular $(m_Q^2,\, m_U^2,\, m_{H_U}^2,\, M_a,\, A_t)$, are generated at the high-scale $M$. The value of $m_{H_U}^2$ at the scale $\mathcal Q$ can then be computed on general grounds as
\begin{eqnarray}
m_{H_U}^2(\mathcal Q)&=&m_{H_U}^2+\eta_{Q}[\mathcal Q,M](m_Q^2+m_U^2+m_{H_U}^2)+\sum_{a}\eta_{a}[\mathcal Q,M]M_a^2\nonumber\\
&+&\sum_{a\neq b}\eta_{ab}[\mathcal Q,M]M_aM_b+\sum_a \eta_{aA}[\mathcal Q,M]M_a A_t+\eta_{A}[\mathcal Q,M]A_t^2 + \Delta_{Y,H_U}
\label{analytic}
\end{eqnarray}
where the soft breaking terms on the right-hand side are defined at the scale $M$ and the coefficients $\eta_X$ depend on the scales $M$ and $\mathcal Q$, and the last term represents the hypercharge D-term contribution.   This expression describes the one-loop evolution of the 
Higgs square mass parameter $m_{H_U}^2$,  when the bottom and tau Yukawa couplings are small, as happens for moderate values of $\tan\beta$. The hypercharge $D$-term vanishes in all the supersymmetry breaking schemes that we analyze in this article, and therefore we shall not consider it in our analysis. However, for completeness, in the Appendix we provide the expression of the additional corrections induced by $\Delta_{Y,H_U}$.

\begin{figure}[h!]
\begin{center}
\includegraphics[width=81.9mm]{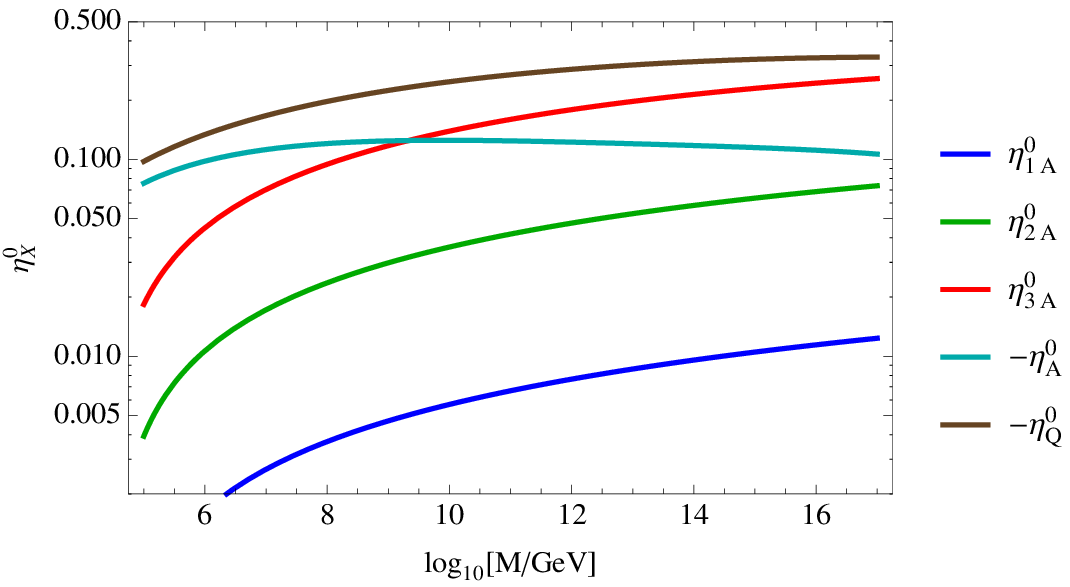}
\includegraphics[width=81.9mm]{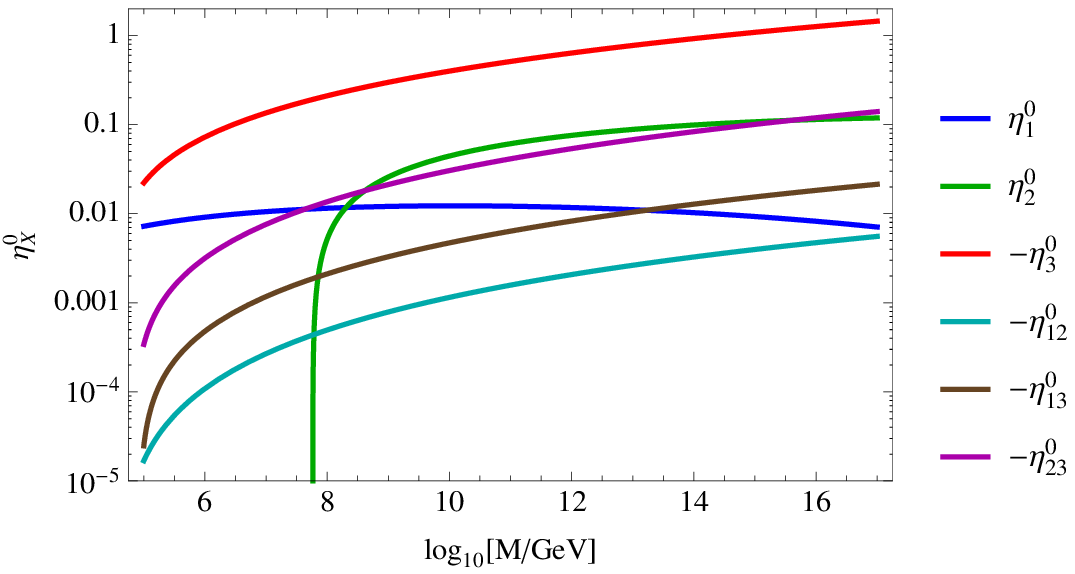}
\end{center}
\caption{\it Left panel: From top to bottom along the left vertical axis, plots of $-\eta_Q^0$, $-\eta_A^0$, $\eta_{3A}^0$, $\eta_{2A}^0$, $\eta_{1A}^0$ as functions of $M$. Right plot: From top to bottom along the left vertical axis, plots of $-\eta_3^0$, $-\eta_1^0$, $-\eta_{23}^0$, $-\eta_{13}^0$, $-\eta_{12}^0$, $-\eta_2^0$ as functions of $M$. }
\label{fig:eta}
\end{figure}

In particular if we fix the low scale as the scale where supersymmetric particles decouple $\mathcal Q=\mathcal Q_0$ (in the few TeV range) we can write the focus point scale as the scale where the condition $m_{H_U}^2(\mathcal Q_0)=0$ is satisfied, i.e.
\begin{eqnarray}
0&=&m_{H_U}^2+\eta_{Q}^0(M)(m_Q^2+m_U^2+m_{H_U}^2)+\sum_{a}\eta_{a}^0(M)M_a^2\nonumber\\
&+&\sum_{a\neq b}\eta_{ab}^0(M)M_aM_b+\sum_a \eta_{aA}^0(M)M_a A_t+\eta_{A}^0(M)A_t^2
\label{FP}
\end{eqnarray}
where now the coefficients $\eta_X^0(M)$ only depend on the (messenger) scale $M$ at which supersymmetry breaking is transmitted. These coefficients are computed semi-analytically~\cite{IL}--\cite{Wagner:1998vd} and their explicit expressions can be found in the Appendix. In fact if we choose a value of $Q_0=2$ TeV they are can be fitted by an expression as
\be
\eta^0_X(M)=\sum_{n\geq 0}a^n_X \, y^n(M),\quad y(M)\equiv\log_{10}(M/GeV)
\label{fitcoefficients}
\ee
where the fitted coefficients are given in Tab.~1 for $\mathcal Q_0=2$ TeV and values of $M$ in the interval $M\in[10^5,10^{17}]$ GeV.
\begin{table}[htb]
\footnotesize
\begin{center}
\begin{tabular}{||c||c|c|c|c|c|c|c|c|c|c|c||}
\hline\hline
$n$&$10\,a_Q^n$&$10^2\,a_1^n$&$10\,a_2^n$&$10^2\,a_3^n$&$10^4\,a_{12}^n$&$10^3\,a_{13}^n$&$10^2\,a_{23}^n$&$10^3\,a_{1A}^n$&$10^2\,a_{2A}^n$&$10\,a_{3A}^n$&$10\,a_A^n$\\
\hline
$0$&1.289&1.124&-2.324&4.276&-5.669&-1.373&-1.022&-4.815&-3.230&-1.329&2.510\\
$1$&-0.529&0.540&0.377&1.820&2.784&0.873&0.598&1.111&0.767&0.333&-1.171\\
$2$&0.015&0.038&-0.010&-0.624&-0.337&-0.121&-0.080&-0.006&-0.009&-0.006&0.134\\
$3$&&0.0007&&&&&&&&&-0.007\\
$4$&&&&&&&&&&&0.0001\\
\hline\hline
\end{tabular}
\end{center}
\label{table1}
\caption{\it Values of the fitted coefficients in (\ref{fitcoefficients}) for $Q_0=2$ TeV}
\end{table}
The functions $\eta_X^0(M)$  are plotted in Fig.~\ref{fig:eta}
from where we can see that for the case of high-scale supersymmetry breaking (i.e.~$M\simeq  10^{16}$ GeV) the plot is dominated by $\eta_3^0$ (i.e.~by the term $M_3^2$) while for low-scale supersymmetry breaking (i.e.~$M\simeq 10^5$ GeV) the function $\eta_Q^0$ (i.e.~the term $m_Q^2+m_U^2+m_{H_U}^2$) takes over and dominates.
Moreover Eq.~(\ref{FP}) is telling us that if we scale all soft-breaking terms as 
\be
(m_Q^2,\, m_U^2,\, m_{H_U}^2,\, M_a,\, A_t)\to\ (\lambda^2\, m_Q^2,\, \lambda^2\, m_U^2,\, \lambda^2\, m_{H_U}^2,\, \lambda\, M_a,\, \lambda\, A_t)
\ee
the equation is still valid. This shows the insensitivity of the FP to the boundary conditions. Alternatively we can leave one of the soft-breaking terms as a free (floating) parameter, e.g.~the boundary value $m_{H_U}^2$. So in general using the scale invariance of Eq.~(\ref{FP}) we can express all masses in units of $m_{H_U} \equiv \sqrt{ |m^2_{H_U}|}$.
Of course the scale at which the FP happens should depend on the boundary values of the soft-breaking terms, so that fixing it to $Q_0$ amounts to a relation between the scale at which supersymmetry is broken $M$ and the soft-breaking terms. 

Let us also stress that the (absolute value of the) mass parameters in the stop sector are constrained by the condition of obtaining the proper value of the observed Higgs mass, $m_H \simeq 125.6$~GeV. This fixes the possible range of values of $Q_0$, which for moderate values of $\tan\beta$  may vary between values lower than a TeV and values of the order of 10~TeV, depending on the value of the stop mixing parameter~\cite{Martin:2007pg}--\cite{Draper:2013oza}.   The general focus point solution define correlations between the different SUSY breaking parameters at the messenger scale $M$ required to make the Higgs mass parameter small at the decoupling scale $Q_0$.  In this article, we have studied these correlations by means of the one-loop renormalization group evolution of the mass parameters. Two loop-corrections, as well as threshold corrections at the scale $Q_0$, will also affect the value of $m_{H_U}^2$.  For large values of the messenger scale $M$, these corrections are expected to be subleading and of the same order as the variation of the Higgs mass parameter in the range of values of $Q_0$ quoted above.  In this article, we have not studied the precise dependence on $Q_0$, but instead we have taken  $Q_0 = 2$~TeV as a representative value. 
Although the Higgs mass parameter can have significant variations with the scale $Q_0$, we don't expect the correlations needed to make $m_{H_U}^2$ small to depend strongly in the precise value of this scale.
In the following section we will explore some popular theories of supersymmetry breaking in where certain relations among the boundary conditions are predicted.

\section{Focus Point for particular models}
\label{models}
As we have seen in the previous section, the FP in the MSSM translates into a condition between the different soft breaking parameters $(m_Q,\, m_U,\, m_{H_U},\, M_1,\, M_2,\, M_3,\, A_t)$ at the scale $M$, the scale at which supersymmetry is transmitted to the observable sector, and hence depends on the precise value of $M$.  In this section we will study the FP predictions in the MSSM assuming some patterns of supersymmetry breaking at the scale $M$, motivated by particularly appealing mechanisms of supersymmetry breaking mediation to the observable sector.

\subsection{The CMSSM}
In this theory [dubbed constrained MSSM (CMSSM)], and inspired by minimal supergravity (SUGRA)~\cite{reviews}, we have as independent parameters $m_0,\, m_{1/2}$ and $A_t$, in such a way that at the scale $M$
\be
m_Q=m_U=m_{H_U}\equiv m_0,\quad M_a\equiv m_{1/2} 
\label{CMSSM}
\ee

\begin{figure}[htb]
\begin{center}
\includegraphics[width=81.9mm]{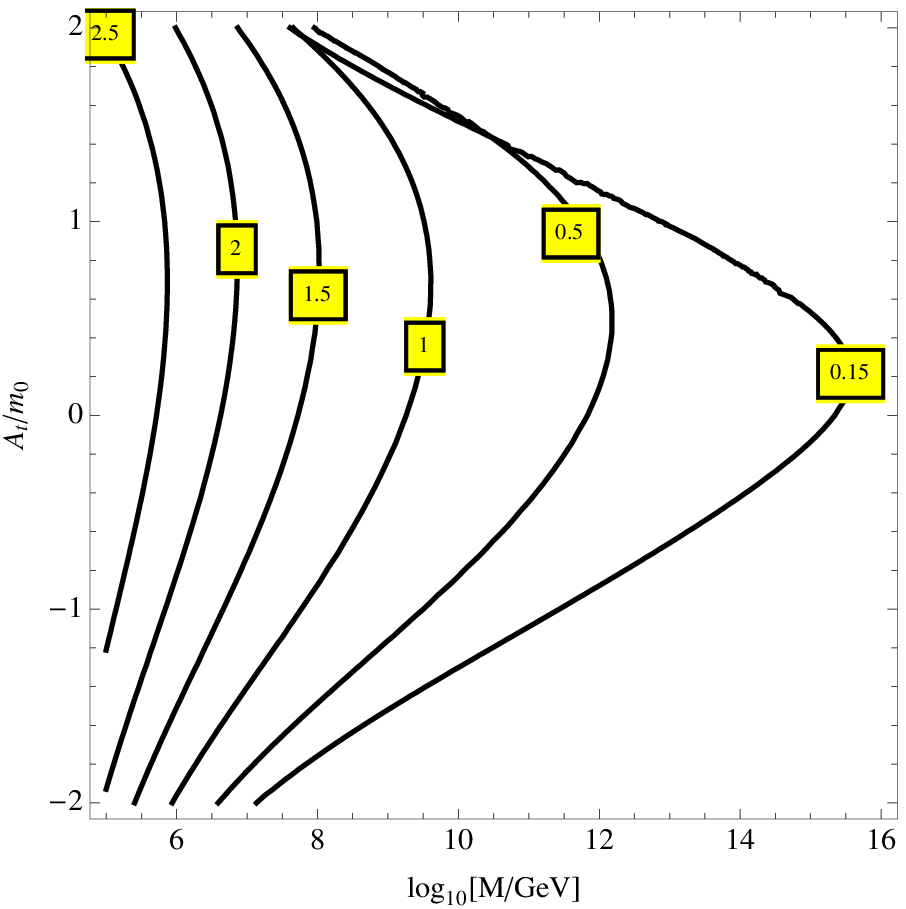}
\includegraphics[width=81.9mm]{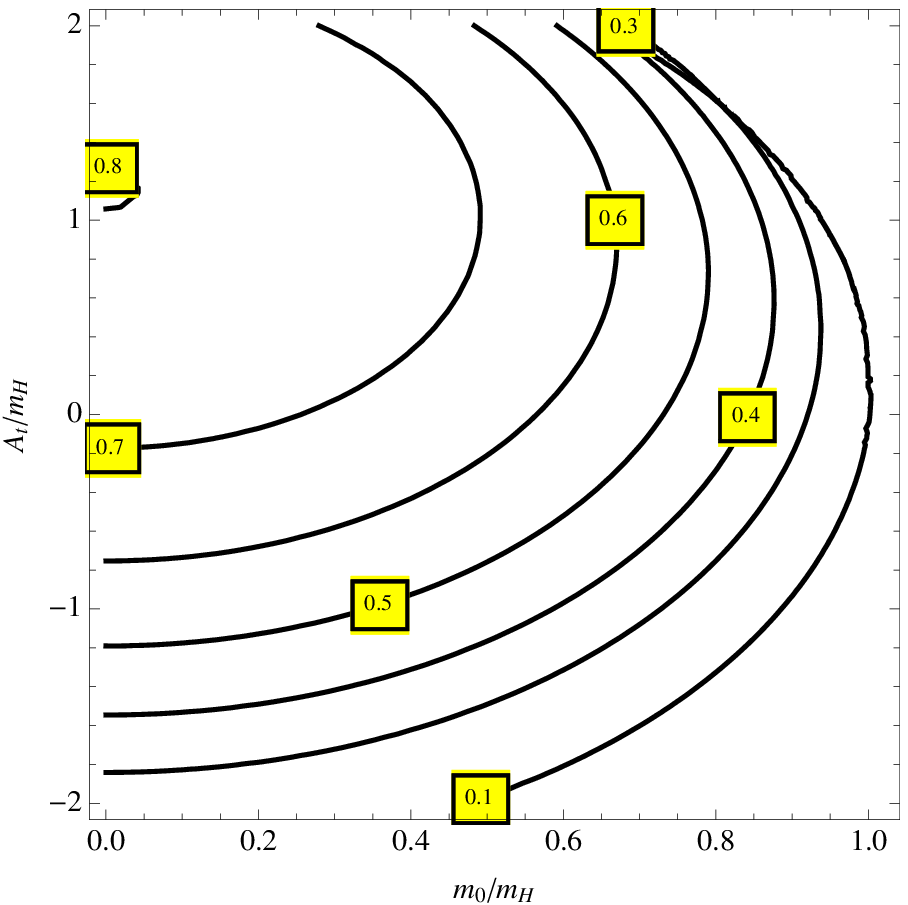}

\end{center}
\caption{\it Left panel: Contour lines of $m_{1/2}/m_0$ in the plane $(\log_{10}[M/\textrm{GeV}],A_t/m_0)$ for the boundary conditions in Eq.~(\ref{CMSSM}). Right panel: Contour lines of of $m_{1/2}/m_H$ in the plane $(m_0/m_H,A_t/m_H)$ for a scale $M=10^{16}$ GeV and the boundary conditions in Eq.~(\ref{HCMSSM}). }
\label{fig:CMSSM}
\end{figure}

We can see from the left panel of Fig.~\ref{fig:CMSSM} that values of $M$ around the unification scale can only be reached for very small values of $A_t$ and $m_{1/2}$.  This is nothing but the original focus point solution found in Ref.~\cite{Feng:1999zg} and is related to an intriguing cancellation of the overall dependence of $m_{H_U}^2$ on $m_0$.   More precisely, the overal coefficient vanishes due to the fact that $\eta_Q \simeq 1/3$.  Since, from Eq.~(\ref{mQ2}), $\eta_Q = y/2$, this happens because at large $\tan\beta$ the square of the Yukawa coupling is close to two thirds of the value that it would obtain if it were very large at the scale $M \simeq 10^{16}$~GeV.  For smaller values of $M$ one sees that the value of $m_{1/2}$ in general increases and that solutions with large $A_t$ exist.

 
 One can think that the fine-tuning will then be eliminated when $m_{H_U}^2 \simeq 0$, since when one varies all parameters at once, respecting the given correlation, the value of $m_{H_U}^2$ remains small.  Of course, as we already mentioned,  that will only be true if the correlation among the different parameters is indeed a prediction of the UV theory.

The correlation between scalar and gaugino masses changes if one does not assume universality of scalar and/or gaugino masses. So a variant of the CMSSM where universality of scalar masses is given up, often considered in the literature and motivated by string constructions where the Higgs fields live in a different location to the rest of matter fields~\cite{Ibanez:1992hc}, has an extra parameter $m_H$. It is dubbed NUHM1~\cite{AbdusSalam:2011fc} and the soft-breaking parameters are then~\footnote{One such model has been constructed in Ref.~\cite{oai:arXiv.org:1201.5164} to provide, by the RGE running, the light stop scenario and further motivated by electroweak baryogenesis studies in the MSSM.}
\be
m_Q=m_U\equiv m_0,\quad m_{H_U}=m_{H_D}\equiv m_H,\quad M_a\equiv m_{1/2}
\label{HCMSSM}
\ee
For such a model contour lines of $m_{1/2}/m_H$ for $M=10^{16}$ GeV are shown in the right panel of Fig.~\ref{fig:CMSSM} where we see that solutions where all masses are heavy do exist.  Larger values of $m_{H_U}/m_0$ generate a positive  coefficient in $m_0^2$, while larger values of $A_t$ do the opposite. The overall coefficient in $m_{1/2}^2$ remains negative.  Hence, solutions exist only when $m_0/m_{H_U} < 1$ and large values of $A_t$ can only be obtained for small values of the gaugino masses. The asymmetry between positive and negative values of $A_t$ observed in this figure is due to the existence of a non-vanishing coefficient in $m_{1/2} A_t$.

\begin{figure}[htb]
\begin{center}
\includegraphics[width=81.9mm]{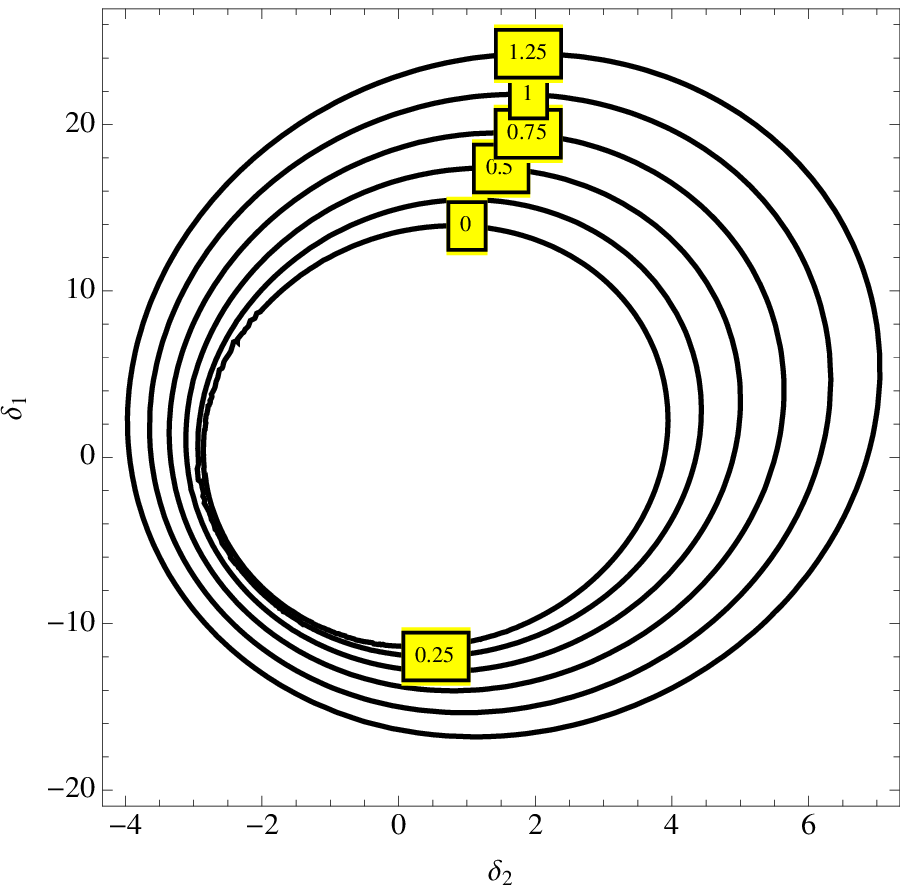}
\includegraphics[width=81.9mm]{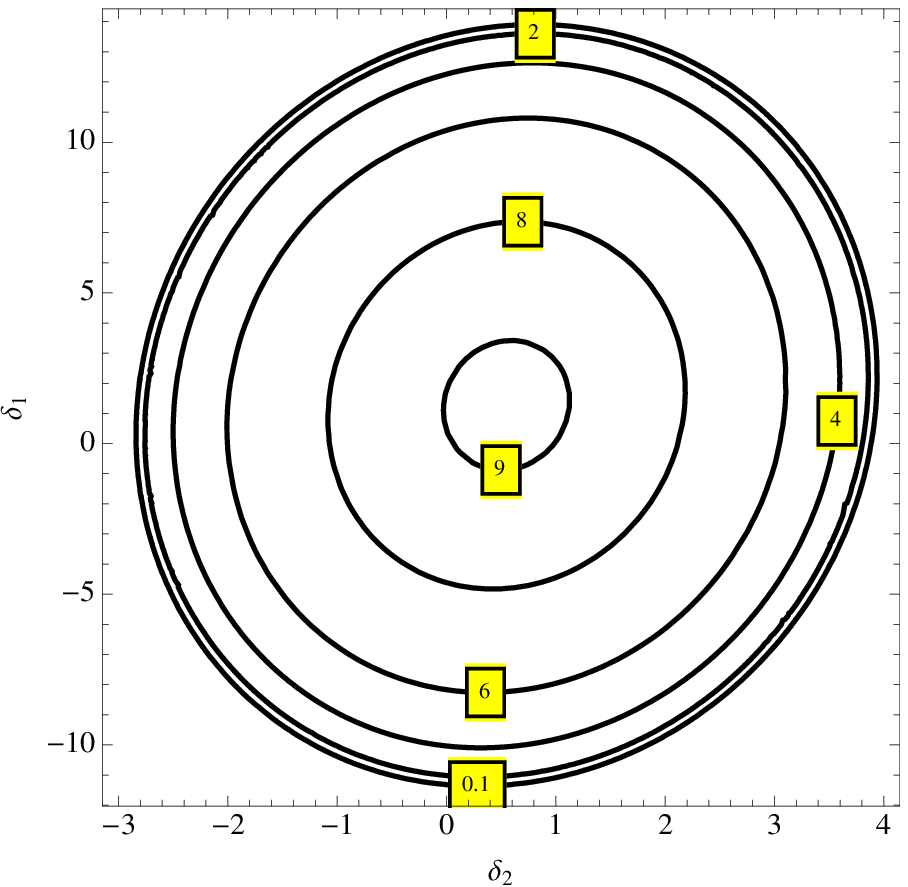}

\end{center}
\caption{\it Left panel: Contour lines of $m_0/m_{1/2}$ in the plane $(\delta_2,\delta_1)$ for $M=10^{16}$ GeV and $A_t=-2.5 m_0$. Right panel: The same for $A_t=0$. }
\label{fig:gaugino}
\end{figure}

Another possibility is giving up  universality of the gaugino masses. To this end we will introduce extra parameters $\delta_a$ such that
\be
M_a=\delta_a m_{1/2}
\ee
where one of the parameters can be fixed to one as it can be reabsorbed in a redefinition of the parameter $m_{1/2}$. A nontrivial pattern for the $\delta_a$-parameters can arise in the effective theories of string constructions~\cite{Brignole:1993dj}. In particular an analysis of fine-tuning has been done in Ref.~\cite{Kaminska:2013mya} where $\delta_3=1$ has been fixed.  In this case the contours lines of $m_0/m_{1/2}$ for $M=10^{16}$ GeV are shown in Fig.~\ref{fig:gaugino} as a function of $\delta_1$ and $\delta_2$ for $A_t=-2.5m_0$ (left panel) and $A_t=0$ (right panel).
In the left panel, for $A_t=-2.5 m_0$, the region where gauginos are heavy, i.e.~$m_0/m_{1/2}\in [0,1]$, is much larger than in the right panel, for $A_t=0$, which corresponds to the external ring.  The reason for the change in the position of the ellipses for which the fine-tuning disappears is again the fact that for $A_t = 0$ the overall coefficient in $m_0^2$ is small and positive and therefore it can be cancelled when the overall coefficient on $m_{1/2}^2$ is negative.  On the contrary, for non-vanishing $A_t = -2.5 m_0$, the overall coefficient on $m_0$ is negative and can only be cancelled when the overall coefficient on $m_{1/2}^2$ becomes positive.
The contour  $m_0/m_{1/2} = 0$ corresponds to the condition under which the overall coefficient controlling the  $m_{1/2}^2$ dependence of $m_{H_U}^2$ vanishes, and it is the same, independently of the relation of $A_t$ with $m_0$, as can be easily seen by comparing the left and right panels of Fig.~\ref{fig:gaugino}.   When such condition is approximately fulfilled, the dependence on both $m_0^2$ and $m_{1/2}^2$ becomes small, and then the fine-tuning is greatly reduced, a result that was first pointed out in Ref.~\cite{Kaminska:2013mya}.

\subsection{Gauge mediation}

In gauge mediation models~\cite{Giudice:1998bp}, supersymmetry is broken in a hidden sector $(X=M+\theta^2 F)$ coupled to a number of messenger fields (charged under the standard model gauge groups) by the following superpotential coupling $W=\Phi^I X\bar\Phi_I$. Supersymmetry breaking is then communicated to the visible sector via gauge interactions generating the following soft breaking masses:
\begin{align}
m_Q^2&=2\left(\frac{4}{3}\alpha_3^2+\frac{3}{4}\alpha_2^2+ \frac{1}{60}\alpha_1^2 \right) \Lambda_S^2\nonumber\\
m_U^2&=2\left(\frac{4}{3}\alpha_3^2+ \frac{4}{15}\alpha_1^2 \right)\Lambda_S^2\nonumber\\
m_{H_U}^2&=2\left(\frac{3}{4}\alpha_2^2+ \frac{3}{20}\alpha_1^2 \right)\Lambda_S^2\nonumber\\
M_a&=\alpha_a  \Lambda_G,\quad A_t=0
\label{GM}
\end{align}
where  in minimal models $\Lambda_G=NF/4\pi M$ and $\Lambda_S=\sqrt{N} F/4\pi M$, $N$ being the number of messengers. Vanishing values of the stop mixing parameter at the messenger scale imply that in order to obtain the proper Higgs mass one needs either large values of the messenger scale or a heavy supersymmetric spectrum~\cite{Patrick}. 

\begin{figure}[htb]
\begin{center}
\includegraphics[width=90.mm]{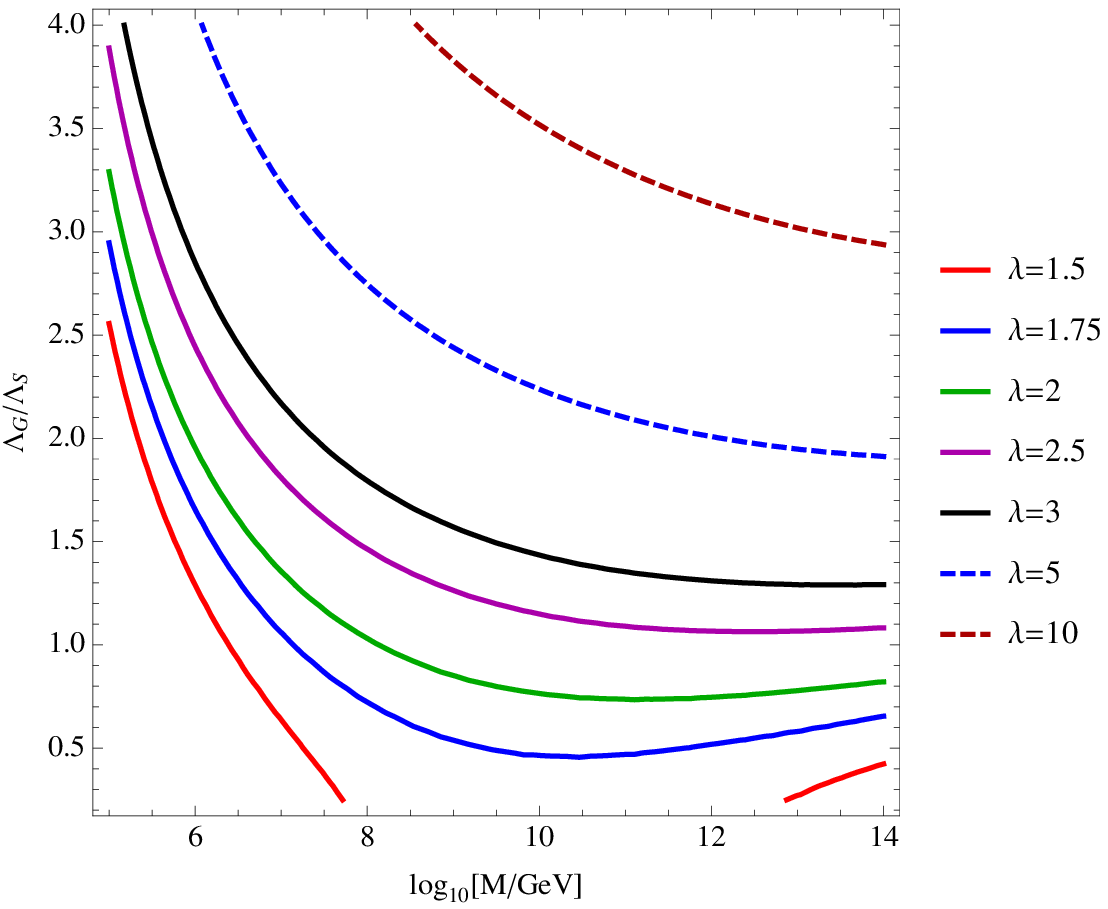}
\includegraphics[width=73.7mm]{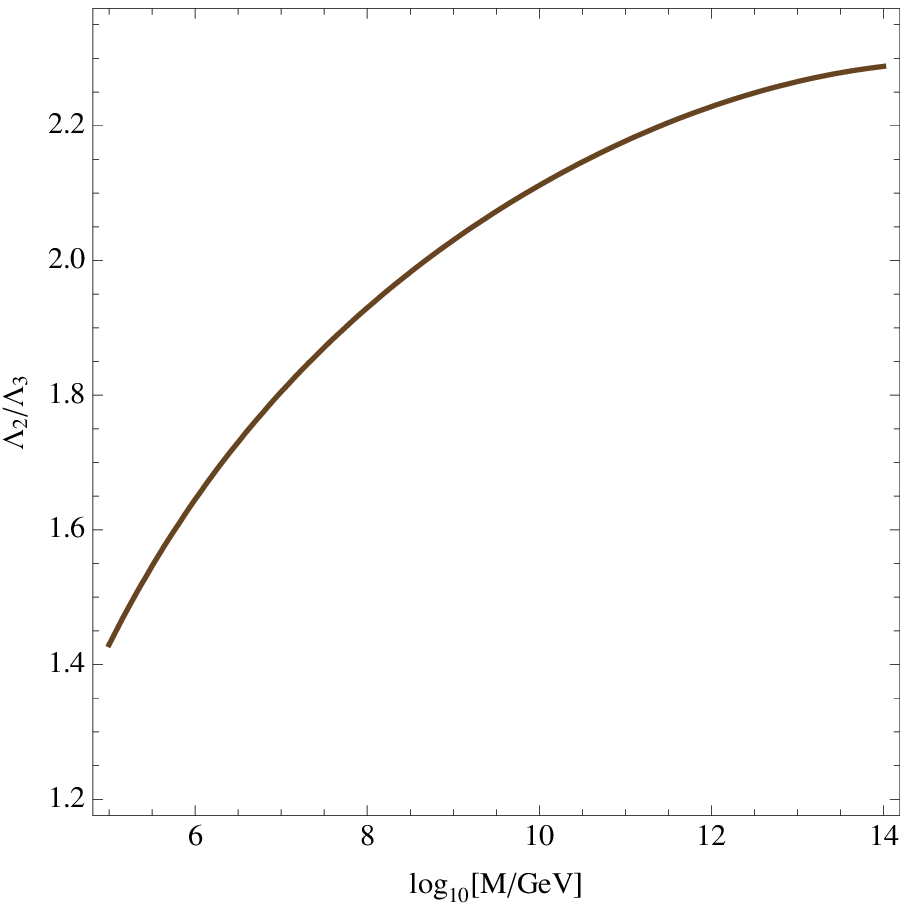}
\end{center}
\caption{\it Left panel: FP lines for constant $\lambda$ in the $(\log_{10}[M/\textrm{GeV}],\Lambda_G/\Lambda_S)$ plane in the model defined by Eqs.~(\ref{GM}) and (\ref{masatotal}). Right panel: FP lines in the plane $(\log_{10}[M/\textrm{GeV}],\Lambda_2/\Lambda_3)$ for the model of Eq.~(\ref{carlos}).}
\label{fig:GGM}
\end{figure}

In order to explain the generation of the $\mu/B_\mu$ terms in gauge mediation we should admit direct couplings with hidden sector operators, as in the superpotential $W=\lambda_U H_U \mathcal O_D+\lambda_D H_D \mathcal O_U$~\cite{Komargodski:2008ax,DeSimone:2011va}, which lead at one-loop to the masses
\be
\delta m^2_{H_{U,D}}=|\lambda_{U,D}|^2\Lambda_{D,U}^2
\ee
where $4\pi\Lambda_{D,U}$ parametrizes the contributions coming from the two-point function of
the $F$-component of the hidden sector operators $\mathcal O_{D,U}$. Assuming then that $\Lambda_{U,D}\simeq \Lambda_S$ the electroweak symmetry breaking requirement leads to values $\lambda_{U,D}\sim \alpha_2$ in which case the total contribution to the Higgs mass can be parametrized as 
\be
m_{H_{U,D}}^2+\delta m_{H_{U,D}}^2\equiv (1+\lambda)\ m_{L}^2
\label{masatotal}
\ee
where $L$ is the slepton doublet and the parameter $\lambda$, $0\lesssim \lambda \lesssim \mathcal O$(few), parametrizes the correction to $m_{H_U}^2$ in Eq.~(\ref{masatotal}).  
In the left panel of Fig.~\ref{fig:GGM} 
we show contour lines of constant $\lambda$ in the plane $(\log_{10}[M/\textrm{GeV}],\Lambda_G/\Lambda_S)$. We can see that there is no FP for the case $\lambda=0$ which corresponds to the case of standard gauge mediation.  The behavior is easily understood due to the fact that for $\lambda = 0$, the overall coefficient on $m_{H_U}^2$ becomes negative, and the overall coefficient on the overall gaugino masses is also negative.  When $\lambda$ is sizable, instead, the overall coefficient on $m_{H_U}^2$ becomes positive and then there exist correlations for which $m_{H_U}^2$ vanishes. 

A variant of the previous models happens when there are several fields $X_I$  in the hidden sector such that the coupling with the messengers is by the superpotential $W=\Phi^I X_I \bar\Phi_I$. In this case different messenger components $X_I$ are affected by different breakings $F_I$. A simple model along these lines was constructed in Refs.~\cite{Wagner:1998vd},~\cite{Martin:1996zb}, where there is a pair of messengers in the $\textbf{5}+\overline{\textbf{5}}$ representation of $SU(5)$  which decompose into the $SU(3)$ triplet ($I=3$) and $SU(2)$  doublet ($I=2$) components. Correspondingly
there are two fields in the hidden sector: $X_3=M+\theta^2 F_3$ giving a mass to the gluino and colored scalars and $X_2=M+\theta^2 F_2$ whose auxiliary component $F_2$ gives a mass to the $SU(2)\otimes U(1)$ gauginos and scalars. The contribution to the soft breakings at the scale $M$ is given by
\begin{align}
m_Q^2&=2\left[\frac{4}{3}\alpha_3^2 \Lambda_3^2+\frac{3}{4}\alpha_2^2\Lambda_2^2+
\frac{1}{60}\alpha_1^2\left(\frac{2}{5}\Lambda_3^2+\frac{3}{5} \Lambda_2^2 \right)   \right]\nonumber\\
m_U^2&=2\left[\frac{4}{3}\alpha_3^2 \Lambda_3^2+
\frac{4}{15}\alpha_1^2\left(\frac{2}{5}\Lambda_3^2+\frac{3}{5} \Lambda_2^2 \right)   \right]\nonumber\\
m_{H_U}^2&=2\left[\frac{3}{4}\alpha_2^2 \Lambda_2^2+
\frac{3}{20}\alpha_1^2\left(\frac{2}{5}\Lambda_3^2+\frac{3}{5} \Lambda_2^2 \right)   \right]\nonumber\\
M_1&=\alpha_1 \left(\frac{2}{5}\Lambda_3+\frac{3}{5}\Lambda_2 \right)\nonumber\\
M_2&=\alpha_2\Lambda_2,\quad M_3=\alpha_3\Lambda_3
\label{carlos}
\end{align}
where $\Lambda_I=F_I/4\pi M$. The FP results for this model are shown in the right panel of Fig.~\ref{fig:GGM}.  Large values of $\Lambda_2/\Lambda_3$ reduces the stop contributions and lowers the negative dependence on the overall scalar mass parameters, implying the existence of solutions. Since all parameters are correlated, for each scale $M$, the solutions occur for a specific value of the ratio $\Lambda_2/\Lambda_3$.

\subsection{Mirage mediation}
\begin{figure}[htb]
\begin{center}
\includegraphics[width=120.mm]{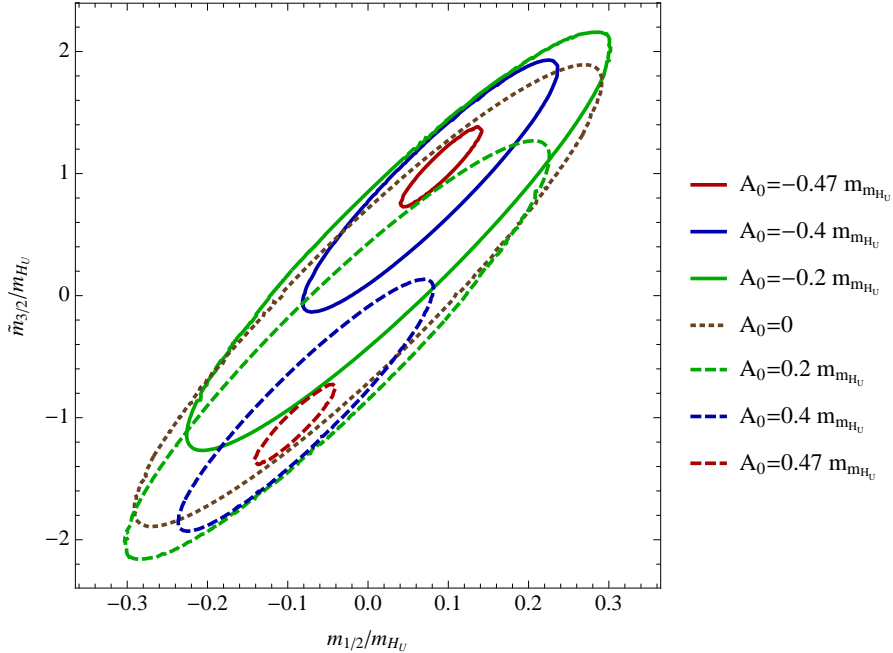}
\end{center}
\caption{\it Contour lines of constant $A_0$ in the plane $(m_{1/2}/m_{H_U},\widetilde m_{3/2}/m_{H_U})$ for a fixed value of the high scale, $M=10^{16}$ GeV.}
\label{fig:mirage}
\end{figure}

This scenario is inspired by string compactification with fluxes~\cite{Kachru:2003aw} and it is dubbed as mixed-modulus-anomaly mediated supersymmetry breaking and also mirage mediation~\cite{Choi:2004sx,Choi:2005ge} as gaugino masses "apparently" unify at a scale much below $M_{\rm GUT}$. Mirage mediation assumes that the contributions from gravity mediation~\cite{reviews} and anomaly mediation~\cite{Randall:1998uk} are comparable in size. Anomaly mediation assumes that supersymetry breaking is communicated via the trace anomaly of any non-conformal theory and it is proportional to the RGE evolution of parameters. Therefore the spectrum at the supersymmetry breaking scale $M$ is given by
\begin{align}
m_{H_U}^2&=m_0^2+\left[3\alpha_t\left(6\alpha_t-\frac{16}{3}\alpha_3-3\alpha_2-\frac{13}{15}\alpha_1\right)-\frac{3}{2}\alpha_2^2b_2-\frac{3}{10}\alpha_1^2 b_1\right]\widetilde m_{3/2}^2\nonumber\\
m_{Q}^2&=m_0^2+\left[\alpha_t\left(6\alpha_t-\frac{16}{3}\alpha_3-3\alpha_2-\frac{13}{15}\alpha_1\right)-\frac{8}{3}\alpha_3^2 b_3-\frac{3}{2}\alpha_2^2b_2-\frac{1}{30}\alpha_1^2 b_1\right]\widetilde m_{3/2}^2\nonumber\\
m_{U}^2&=m_0^2+\left[2\alpha_t\left(6\alpha_t-\frac{16}{3}\alpha_3-3\alpha_2-\frac{13}{15}\alpha_1\right)-\frac{8}{3}\alpha_3^2 b_3-\frac{8}{15}\alpha_1^2 b_1\right]\widetilde m_{3/2}^2\nonumber\\
A_t&=A_0-\left(6\alpha_t-\frac{16}{3}\alpha_3-3\alpha_2-\frac{13}{15}\alpha_1\right)\widetilde m_{3/2}\nonumber\\
M_a&=m_{1/2}+\alpha_a b_a \widetilde m_{3/2}
\end{align}
where $\widetilde m_{3/2}=m_{3/2}/4\pi$ and $b_a=(33/5,1,-3)$. It is known that for $m_0 = 0$, the slepton square masses become negative and therefore no physical solution exist~\cite{Randall:1998uk}. For positive values of $m_0$ and $m_{1/2}$ instead,
positive slepton masses may be obtained. It is easy to show that whenever $m_{H_U}(Q_0) = 0$, the slepton masses become positive at the same scale.  It is also easy to show that in order to find such a solution, the value of $m_{H_U}^2$ at the scale $M$ must be positive. 
In Fig.~\ref{fig:mirage} we plot in the plane $(\widetilde m_{3/2},m_{1/2})$ contour lines of constant value of $A_0$ for $M=10^{16}$ GeV. Here all masses are normalized to the value of $m_{H_U} \equiv \sqrt{m_{H_U}^2}$ at the scale $M$. The solutions are symmetric under a simultaneous change of sign of  $m_{1/2}$, $\tilde{m}_{3/2}$ and $A_0$. Moreover, solutions may be only obtained for moderate values of $A_0/m_{H_U}$ and disappear for large values of this parameter. 

\section{Conclusion}
\label{conclusion}

In this article we have found the conditions to obtain small values of the  soft supersymmetry breaking parameter of the Higgs field
at low energies, even in the case where the high energy values of the scalar and gaugino supersymmetry breaking parameters are much larger
than the weak scale.  These conditions do not depend on the overall scale of the supersymmetry breaking parameters, and therefore
define correlations between the different mass parameters.  When these conditions are fulfilled, the proper electroweak symmetry 
breaking may be obtained for small values of $\mu$, without the need of fine tuning.  Needless to say, this fine tuning is traded for the
above defined correlations, which, however, may have a dynamical origin. 

For universal values of the scalar mass parameters, the conditions are fulfilled for a large messenger scale, of the order of the GUT scale,
moderate or large values of $\tan\beta$ (but not large enough to make the bottom Yukawa effects relevant)  and for values of the gaugino masses 
significantly smaller than the scalar masses. This is nothing but the original Focus Point solution and therefore we have defined the solutions found 
in this paper as General Focus Point solutions.  Using the one-loop renormalization group evolution of the mass parameters,  our article defines 
the correlations that the soft supersymmetry breaking parameters must fulfill, for all reasonable values of the messenger scale and
for arbitrary high energy values of these parameters. 

We also analyzed particular cases in which the supersymmetry breaking mechanism is such that there is an automatic correlation between some
of the soft supersymmetry breaking parameters. This include the case of universal scalar masses, non-universal Higgs mass parameters, non-minimal
gauge mediation and the mirage mediation extension of the anomaly mediation mechanism. In all cases, the requirement of a small Higgs mass
parameter leads to additional correlations beyond those that are implicit in the given mechanism.   In order to really improve the fine tuning
problem, these correlations should have a dynamical origin, and would lead naturally to a small weak scale. 

\section*{Acknowledgments}
The authors would like to thank the KITP in Santa Barbara for its hospitality at the early stages of this work and support under the National Science Foundation grant PHY-1125915. The work of AD was partially supported by the National Science Foundation under grant  PHY-1215979. The work of MQ was supported in part by the European Commission under the ERC Advanced Grant BSMOXFORD 228169, by the Spanish Consolider-Ingenio 2010 Programme CPAN (CSD2007-00042), and by CICYT-FEDER-FPA2011-25948. The work of CW at ANL is supported in part by the U.S. Department of Energy under Contract No. DE-AC02-06CH11357.

\appendix

\section{General Focus Point Equations}
\label{appendix}

In this appendix we are going to summarize the analytical solution to the RGE evolution of the soft masses the leads to Eq.~(\ref{analytic}), following 
the results derived in Refs.~\cite{IL},\cite{Barbieri:1987fn},\cite{COPW},\cite{Wagner:1998vd}.
Let us start with the solution to the top-quark Yukawa coupling. In general, we can encode the top-quark Yukawa coupling dependence on the scale in
terms of a variable $y$, defined as the
ratio of the square of the Yukawa coupling at the weak scale over the value that  would be obtained if the Yukawa coupling at
the messenger scale $M$ were large,
\begin{equation}
y = - \frac{ 6 \alpha_t(M) F(\mathcal Q)}
{(4\pi) \left[ 1 - \frac{ 6 \alpha_t(M) F(\mathcal Q)}{4\pi}\right]} ,
\end{equation}
where $\mathcal Q<M $ and $\alpha_t = h_t^2/(4 \pi)$.
We will ignore the bottom Yukawa coupling,
considering it to be small compared with the top Yukawa coupling. 
The effect of the bottom Yukawa coupling  on the evolution of the soft
supersymmetry breaking parameters may be  relevant if 
$\tan\beta$ is very large, $\tan\beta \gtrsim m_t/m_b$.
Hence, our analysis is valid in the small and moderate $\tan\beta$ regimes.

The function $F$ is defined by
\begin{equation}
F(\mathcal Q)
= \int_{M}^{\mathcal Q} E(t) dt, \;\;\;\;\;\;\;\;\;
E(\mathcal Q) = \prod_a \left( \frac{\alpha_a(\mathcal Q)}{\alpha_a(M)}
\right)^{-c_t^a/b_a} ,
\end{equation}
where $t = \log(\mathcal Q^2)$,  $b_a$ is the $\beta_a$-function coefficient 
of the gauge group $G_a$ in the effective theory defined in the energy
range between the scales $M$ and $\mathcal Q$, and $c_t^a = 2 (c_{Q_L}^a + c_{U_R}^a + c_{H_U}^a)$ with
$c_Q^a$ being the quadratic Casimir of the superfield 
$Q$ under the $G_a$ gauge group, which, for a fundamental representation
of $SU(N)$ takes the value $c_Q = (N^2-1)/2N$, while $c_Q =  3/5 \times 
(Q_Q - T_3)^2$ for $U(1)$, where $Q_Q$ is the charge of the field $Q$. We are implicitly working 
with a normalization of the gauge couplings consistent with their unification  at high energies, so
$\alpha_1 = 5/3 \; \alpha_Y^{SM}$. 
We will also define 
the linear integral function $H$  by~\cite{COPW},
\begin{equation}
H(f) = \int_{M}^{\mathcal Q} f(t) dt - 
\frac{1}{F(\mathcal Q)}
\int_{M}^{\mathcal Q} F(t) f(t) dt,
\end{equation}
which is most useful to express the soft supersymmetry-breaking
parameters at the weak scale in a compact form.

The soft supersymmetry breaking parameters at low energies may be obtained in terms of their
 boundary conditions at the messenger scale and functions that depend on the gauge and
 Yukawa couplings of the theory.  The gaugino masses have a very simple dependence at
 one-loop, namely 
\begin{eqnarray}
M_a(\mathcal Q) & = & M_a(M) \frac{\alpha_a(\mathcal Q)}
{\alpha_a(M)}.
\end{eqnarray}
The trilinear soft supersymmetry-breaking mass terms, instead,
not only depend on the gauge sector, but are also affected
by top quark Yukawa dependent effects, that modify  their
renormalization group evolution. \begin{eqnarray}
A_t(\mathcal Q) & = & 
\frac{c_t^a M_a(M) }{4 \pi \alpha_a(M)}
\left[ \int_{M}^{\mathcal Q} \alpha_a^2(t) dt - y H(\alpha_a^2)
\right]
\nonumber\\
& + & A_t(M) \left( 1 - y \right) \; . 
\label{atev}
\end{eqnarray}
The scalar mass parameters are also affected by gauge and 
top-quark Yukawa
contributions. The Yukawa dependence enters only through the
renormalization group evolution of these parameters. 
In the case of real gaugino masses, the general
expression is given by
\begin{eqnarray}
\label{mQ2}
m_Q^2(\mathcal Q) & = & m_Q^2(M) 
- 4 c_Q^a \frac{M_a^2(M)}{\alpha_i^2(M)} 
\times \int_{M}^{\mathcal Q} \frac{\alpha_a^3(t)}{4\pi} dt
\nonumber\\
& - &\frac{y d_Q^t}{6} \left(m_{Q_L}^2(M) + m_{U_R}^2(M)
+ m_{H_U}^2(M) \right)
\nonumber\\ 
& - & \frac{2 d_Q^t y}{6} \frac{c_t^b M_b(M) 
c_t^a M_a(M)}
{\alpha_a(M) \alpha_b(M)} 
H\left( \frac{\alpha_b^2(t)}{(4\pi)^2} \int_{M}^{\mathcal Q}
\alpha_a^2(t') dt' \right)
\nonumber\\
& + & \frac{2 d_Q^t y}{6} 
\frac{c_t^a M_a^2(M)}{\alpha_a^2(M)}
H\left(\frac{\alpha_a^3}{4\pi}\right)
+ \frac{d_Q^t}{6} \left(\frac{y 
c_t^a M_a(M)}{\alpha_a(M)}
H\left(\frac{\alpha_a^2}{4\pi}\right) \right)^2
\nonumber\\
& - & \frac{d_Q^t y ( 1- y)}{6} 
\frac{2 c_t^a M_a(M)}{\alpha_i(M)}
H\left(\frac{\alpha_a^2}{4\pi}\right) A_t(M)
\nonumber\\
& - & \frac{d_Q^t y ( 1 - y)}{6} A_t(M)^2  + \Delta_{Y,Q}(\mathcal Q)
\end{eqnarray}

\noindent
where $d_{Q_L}^t = 1$, $d_{U_R}^t = 2$, and $d_{H_U}^t = 3$, where $Q_L$ and $U_R$  are the third
generation left-handed quark doublet and right-handed up quark singlet,
respectively, and $H_U$ is the Higgs that couples to up-type quarks superfields. This expression for the case of $H_U$ corresponds to Eq.~(\ref{analytic}).
The term $\Delta_Y$ comes from the possible contribution of the hypercharge $D$-terms, which takes the generic form
\begin{equation}
\Delta_{Y,Q}(\mathcal Q) = \frac{1}{11} Y_Q \times  \frac{g_1^2(\mathcal{Q}) - g_1^2(M)}{g_1^2(M)} \times \sum_Q Y_Q m^2_Q(M)
\end{equation}
where $Y_Q =  Q_Q - T_3$ is the SM hypercharge and
\begin{equation}
\sum_Q Y_Q m^2_Q   = m_{H_U}^2 - m_{H_D}^2 + \sum_i \left( m_Q^2 - 2 m_U^2 + m_D^2 -  m_L^2 + m_E^2 \right)_i
\end{equation}
and $i$ is a generation index.  It is easy to prove from here that $\sum Y_Q m^2_Q/g_1^2$ is a renormalization group invariant quantity~\footnote{There are 14 RG invariants
in the MSSM, that can be efficiently used to extract information of the underlying theory if sparticles were observed~\cite{RGinv}.}.   The dependence of $\Delta_{Y,Q}$ on the
gauge couplings may be rewritten taking into account that
\begin{eqnarray}
\frac{1}{\alpha_1({\mathcal Q})} & = & \frac{1}{\alpha_1(M_t)} - \frac{41}{20 \pi} \log\left(\frac{\mathcal Q}{M_t}\right)
\nonumber\\
\frac{1}{\alpha_1(M)} & = & \frac{1}{\alpha_1({\mathcal Q})} - \frac{33}{10 \pi} \log\left(\frac{M}{\mathcal Q}\right)
\end{eqnarray}
and $M_t$ is the value of the top mass.  Hence, for the particular case of $H_U$
\begin{equation}
\Delta_{Y,H_U} = - \frac{1}{22} \times \frac{ 66 \log\left(\frac{M}{\mathcal Q}\right)}{1200 \ \pi  - 41 \log\left(\frac{\mathcal Q}{M_t} \right)}  \sum_Q Y_Q m_Q^2(M),
\end{equation}
where we have approximated $1/\alpha_1(M_t) \simeq 60$.  The coefficient of $\sum_Q Y_Q m_Q^2(M)$ is small and negative, and its absolute value becomes smaller for lower messenger scales.
For  ${\mathcal Q} = 2$~TeV and the large messenger scale $M = 10^{16}$~GeV, this coefficient is equal to $-0.024$, while for $M = 10^6$~GeV is equal to $-0.005$.  

Finally, let us write the expression for the $B$ parameter,
which governs the relation between the bilinear terms in the
superpotential and the bilinear soft supersymmetry-breaking
parameters in the scalar potential. This is given by
\begin{eqnarray}
B(\mathcal Q) & = & B(M) - \frac{y  d^t_{H_U}}{6}
A_t(M)
\nonumber\\
& + & \frac{4 c_{H_U}^a M_a(M)}{4 \pi \alpha_a(M)} 
\times \int_{M}^{\mathcal Q} \alpha_a^2(t) \; dt
 -\frac{ c_t^a M_a(M)}{4 \pi \alpha_a(M)}
\frac{y d_{H_U}^t}{6} H(\alpha_i^2) .
\end{eqnarray}

 \end{document}